\documentclass{ws-ijmpa}
\usepackage{times}
\usepackage{overcite}

\begin{document}

\markboth{A. D. Erlykin}
         {Models and Phenomenology}

%
\catchline{}{}{}{}{}
%

\title{MODELS AND PHENOMENOLOGY}

\author{\footnotesize A. D. ERLYKIN 
       }

\address{P. N. Lebedev Physical Institute, Leninsky prospect 53,
         Moscow, RU-119991, Russia \\ erlykin@sci.lebedev.ru}

\maketitle

\pub{Received (Day Month Year)}{Revised (Day Month Year)}
 
\begin{abstract}
It is evident that models of the knee should match the observational 
phenomenology. In this talk I discuss a few aspects of phenomenology, which
are important not only for the understanding of the knee origin, but also
for the general problem of the origin of cosmic rays. Among them are the shape of 
the energy spectrum, its irregularity, the sharpness of the knee and its fine 
structure. 
The classification of models is given and some examples of the most recent models are 
discussed. The most probable conclusion deduced from this examination is that the 
knee has an astrophysical origin and the so called 'source' models of the knee are 
most likely among them.    

\end{abstract}

\section{Introduction}

The total number of models pretending to explain the origin of the knee exceeds twenty 
and the number of people who propose these models is about thirty. The number of ideas 
is definitely larger, but I deliberately distinguish {\em ideas} and {\em models}, 
because sometimes the ideas are not developed at all to be considered as a model. The 
quality of the development is also quite different for different models. To the great 
extent it is due to the lack of astrophysical or particle physics data which could 
constrain the models, but there are also some evident subjective reasons. Certainly it 
is our duty to examine as much consequences of the proposed idea
 as we can and to some extent it is our fault that we ignore some evident experimental 
or observational facts which contradict the proposed idea. There is a tradition in 
our P.N.Lebedev Physical Institute that if you propose the idea you should also  
propose the test which can {\bf kill} your idea. We must admit that there is not a lot 
of authors who propose such tests. 

In this talk I examine some features which should be explained by models of 
the knee and requirements which should be satisfied. 

\section{Phenomenology}

Usually we consider just three main characteristics of cosmic rays (CR) which should 
be predicted by the model and checked by the comparison with experimental data - they 
are {\em the energy spectrum, the mass composition and the anisotropy}. Among them 
there might be some phenomenological sub-features, which are sometimes crucially 
important. Quite a good survey of models with their regard to the observed features of 
the energy spectrum and the mass composition is given in \cite{Hoer1}.  
Here I focus only on the problems related to the CR energy spectrum.

Usually we examine such features of the spectrum as the absolute intensity and slopes
below and above the knee. I should like to draw the attention to some other 
sub-features of the spectrum shape viz. its negligible {\em irregularity}, then 
{\em the sharpness and the fine structure} of the knee. 

\subsection{The irregularity}

Since the discovery of extensive air showers (EAS) it has been found that their size 
spectrum has the power law character in wide energy ranges below and above the 
knee. The theory of shock acceleration explained this power law and since then the 
educated guess was that it is supernova explosions which create the shocks, which 
in turn accelerate particles and produce a regular power law of the CR energy 
spectrum. However, we should remark that a stochastic character of supernova explosions
 in space and time prevents the formation of the regular spectrum. The subsequent 
propagation of CR in the interstellar medium (ISM) only emphasizes the 
irregularity since it results in the dominant contribution of nearby and recent 
supernova remnants (SNR) to CR observed at Earth. The accurate Monte-Carlo
realization of the scenario, where only Galactic SNR give birth to observed CR,
shows the extremely irregular character of the spectrum with the very low probability 
to observe the good power law (Figure 1a). 

We consider the irregularity as a very important phenomenological feature and 
propose as the measure of the irregularity the parameter $\Upsilon=\sigma_{\log{I}}$, 
which is the standard deviation from the mean intensity taken in a logarithmic 
scale. As an example, Figure 1b shows the irregularity as the function of primary 
energy for the scenario of standard supernova exploding sporadically in our Galaxy and
producing different spectra, shown in Figure 1a. It is seen that due to the decreasing 
lifetime of higher energy CR in the Galaxy the smaller number of recent SNR 
contributes to the intensity and the irregularity of the spectra increases.        
\begin{figure}

\centerline{
 \epsfig{file=                  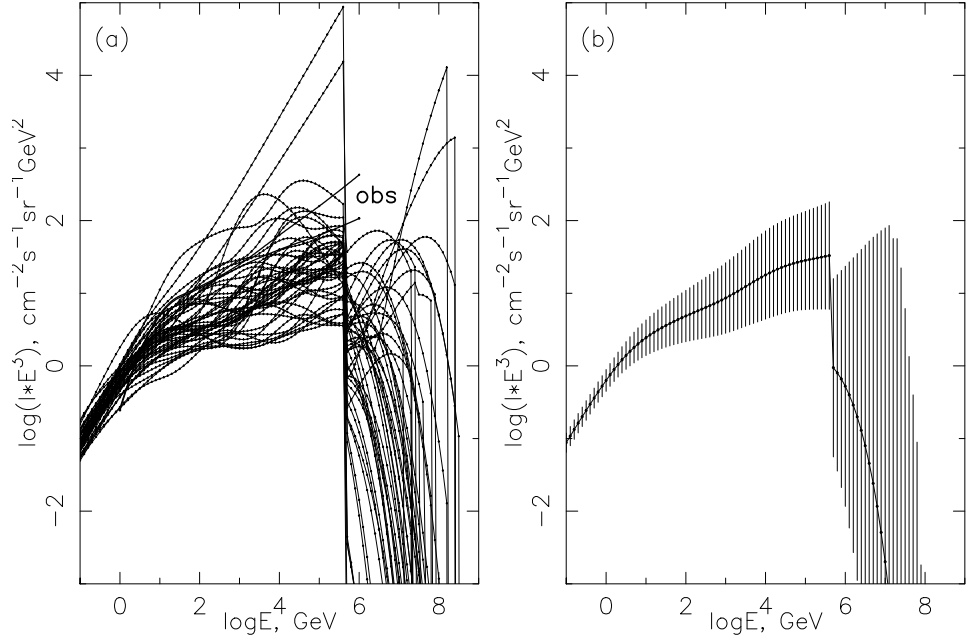,height=\linewidth,angle=-90}
}
\vspace*{8pt}
\caption{(a) A sample of 50 CR energy spectra produced by stochastic explosions of 
supernovae and hypernovae in our Galaxy. Each spectrum is formed by explosions of 49000
 standard supernovae and 1000 hypernovae. (b) The mean spectrum and its irregularity
(standard deviation of the CR intensity). The rise of the irregularity with the 
energy is clearly seen.}
\end{figure}
     
\subsection{The sharpness}  
   
The remarkable sharpness of the knee has been mentioned even in the first publication 
by Kulikov and Khristiansen in 1958 \cite{Kulik}. They have written: {\em 
The presented data indicate with a great probability the possible {\bf sharp} change 
of the spectrum in the studied range of shower sizes $\ldots$}. In \cite{EW1} we 
suggested to use as the measure of sharpness the parameter 
$S=-\frac{d^2{\log{I}}}{d(\log{N})^2}$, where $I$ is the shower intensity and $N$ is the 
EAS size. The advantage of this second derivative is that 
it depends {\em neither} on the absolute intensity {\em nor} on 
the slopes of the spectrum below and above the knee. The world survey of 40 spectra 
\cite{EW2} showed that inspite of the big spread of sharpness values nearly all of them
 exceed the value 0.3 expected for the original Galactic Modulation Model (Figure 2).
 This important phenomenological characteristic should be taken into account in every 
reasonable model of the knee and never ignored.
\begin{figure}
\centerline{
 \epsfig{file=                  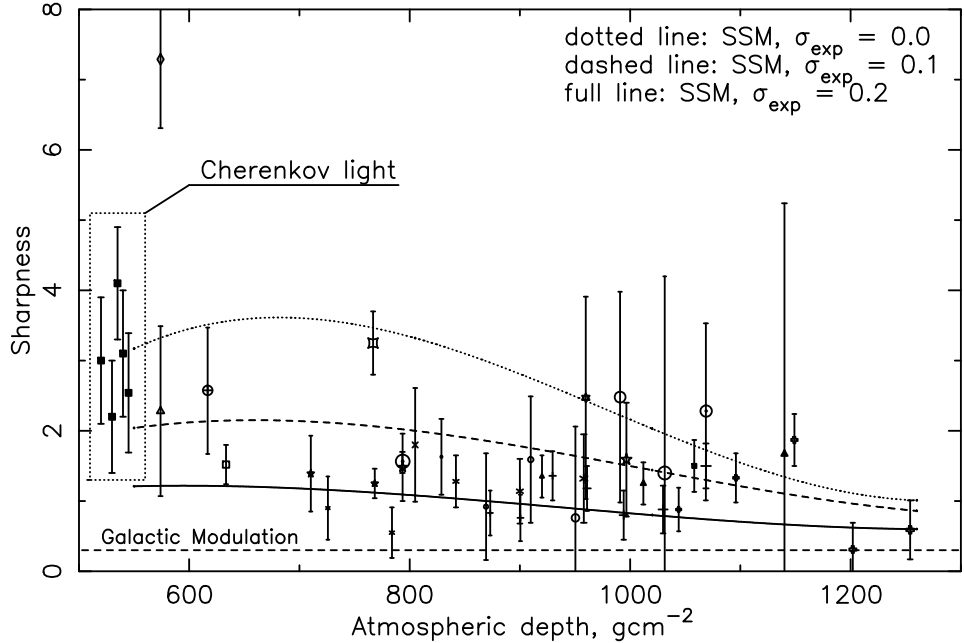,height=\linewidth,angle=-90}
}
\vspace*{8pt}
\caption{The sharpness of the knee in the Cherenkov light (the dashed box) and EAS 
size spectrum vs. the atmospheric depth
of the measurements. The origin of symbols is given in $^4$. The lines are the 
predictions of the Single Source 
Model with an account for the errors of the measurements $\sigma_{\log{N}_e}$:
0 -- dotted line, 0.1 -- dashed line, 0.2 -- full line. The horizontal dashed line at
$S=0.3$ indicates the prediction of the classic Galactic Modulation Model (GMM).
It is seen that the bulk of measured sharpness values exceeds the predictions of the GMM.}    
\end{figure}

\subsection{The fine structure}

Fluctuations in the EAS development smear out the irregularity of the primary CR energy
 spectrum, but not down the regular power law with $\Upsilon = 0$. Due to the 
stochastic nature of the CR sources there {\bf must} be 'structure'{\em at some level}.
 The well known examples of such structures are the widely discussed {\em knee} and 
{\em the ankle}. The search for other structures is difficult since the deviations 
from the power law are evdently small. One must have a good statistics to find the 
structure in the single experiment. To use the data accumulated in many 
experiments one has to eliminate their difference in the background and calibration.
We satisfied the first requirement using
{\em the deviation from the running mean}. To satisfy the second requirement we 
referred the position of the deviation to the position of the knee in each individual
experiment. The position of the knee has been determined as the point of the maximum
sharpness in the spectrum. This approach is absolutely legitimate if you study the 
shape of the spectrum. 

We processed 40 EAS size spectra and 5 Cherenkov light spectra available at that time 
and the result is shown in Figure 3. Besides the apparent excess at the knee there is 
a second excess ('peak') at $\log(E/E_{knee})\approx\log(N_e/N_e^{knee})\approx0.6$ 
seen both in the Cherenkov light and in the EAS electron size spectrum. We consider
this result as the strong evidence for the fine structure of the knee, which should 
be taken into account by any reasonable model of the knee.
\begin{figure}
\centerline{
 \epsfig{file=                  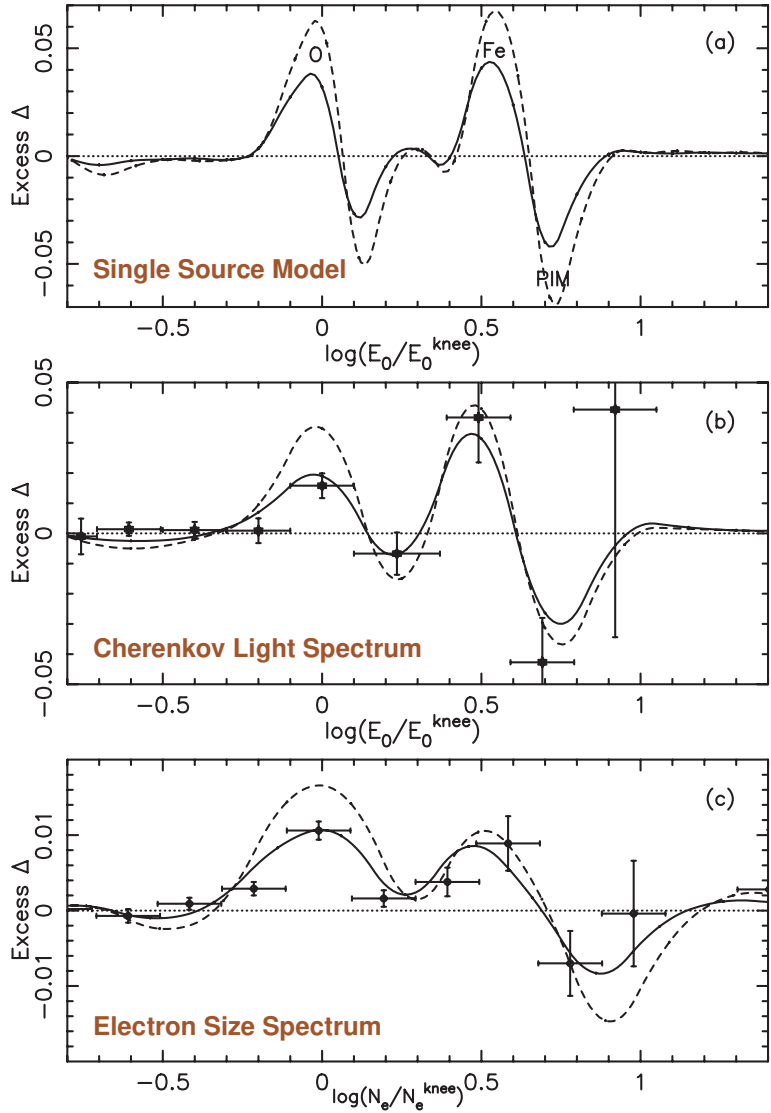,width=0.8\linewidth}
}
\vspace*{8pt}
\caption{The excess over the running mean: (a) in the primary energy spectrum expected 
in the SSM; (b) in the Cherenkov light spectra; (c) in the EAS size spectra. The lines 
indicate the predictions of the SSM for the versions with $O, Fe$ (full line) and 
$He, O$ (dashed line) responsible for the first (knee) and the second peak 
respectively.}
\end{figure}
   
The origin of this peak will be discussed later. However it is clear that such 
structure is nearly impossible to get in the classic Galactic Modulation Model with its
smooth leakage of CR from the Galaxy.

\section{Classification of models}

All models of the knee can be tentatively divided into three groups: 
{\em astrophysical} models, {\em interaction} models and {\em phenomenological} models.
 The astrophysical models in turn can be subdivided into {\em source} models and 
{\em propagation} models. 

The models of the first group explain the origin of the knee by astrophysical reasons: 
either by the change or the evolution of CR sources, or by the change of the 
propagation and the leakage
 from the Galaxy. The interaction models claim that there is no sharp change of the 
primary CR energy spectrum. They refer the steepening of the observed EAS size spectrum
 and the change of their characteristics to the change of the particle interactions at 
PeV energies. The phenomenological models don't specify new sources or new processes 
above the knee, but try to explain the steepening by the existing features of the 
primary CR or an EAS development. Sometimes the association of the model with any group
is ambiguous, and this classification is given just for the orientation. Below I'll 
give some examples of the models, which were discussed in recent years. Certainly the 
list of these examples is not complete and I ask those authors who will not be 
mentioned to apologize me for my ignorance.

\subsection{Astrophysical models}

\subsubsection{Source models}

The most developed conception is that sources of CR below the knee are supernova (SN)
 explosions. The shock waves produced by the explosions can accelerate protons up to 
maximum energies $E_{max} \approx$ 0.1 - 0.4 PeV. The knee is due to the impossibility 
for SN to accelerate particles up to higher energies. This cut-off at $E_{max}$ is 
rather sharp and by this way one can explain the visible sharpness of the knee. 
However the immediate questions to these models are:\\
(i) are SN standard enough to give about the same $E_{max}$ to give the 
sharp knee ? \\
(ii) what is the origin of CR above the knee?

Different source models give different answers to these questions. I should like to
 speak about our Single Source Model later and here just mention the model of 
P.L.Biermann \cite{Bier1}. He allows that there might be just a few ordinary SN 
remnants (SNR) which give dominant contribution at the knee and ensure the formation
of the sharp knee. He even assumes that their explosions due to special conditions 
created during the collapse can be standard \cite{Bier2}. The cosmic rays above the 
knee are produced by another type of SN, which progenitors are more massive, hot and 
emit powerful particle wind before the explosion. The typical representative of such 
type of sources are well known Wolf-Raye stars. The powerful shock wave propagating 
through the bubble created by the hot wind accelerates particle up to EeV energies.

Similar idea is developed by L.G.Sveshnikova \cite{Svesh}, who examined the variety of 
SN types and concluded that the dominant contribution to CR below the knee is given by 
just SNI$_{bc}$ type and above the knee - by explosions of more massive and hot stars
 which she called {\em `hypernovae'} though this name is often associated with SN 
explosions accompanied by gamma-ray bursts (GRB). 

The interesting model of the cosmic ray acceleration by blobs of relativistic plasma is
 proposed by A.Dar, R.Plaga and A.De Rujula \cite{Dar1,Dar2,Plaga,DeRu1,Dar3,DeRu2}. 
Its main subject is the origin of gamma ray bursts, but there are important points in 
this scenario which have relevance to our discussion. The sources of cosmic rays in 
this model are 'cannonballs' - relativistic plasmoids which are emitted by the core 
collapsed supernovae and stopped in the Galactic Halo or in the Intergalactic Medium 
(IGM). By this way authors avoid 
difficulties connected with the need to introduce different sources below and above 
the knee - the sources are the same and the knee is connected with the increasing 
number of reflections from the cannonball, needed to accelerate particles to higher 
energies. Since the model of the knee is not 
sufficiently developed it is difficult to say whether it explains the 
sharpness of the knee or not. As for the irregularity the problem seems not to be 
severe since CR sources can be even more numerous than SNR since each of them can 
emit a few cannonballs. There is no problem with the isotropy and the absence or 
presence of the GZK cutoff, since the sources are mainly distributed in the Galactic 
Halo \cite{Dar2} or in the IGM \cite{Dar3}, but not in the Galactic Disk.        

There are other models which claim that there is no need to introduce another kind of 
sources beyond the knee. SNR are able to accelerate particles up to hundreds PeV due 
to compression and magnification of magnetic fields inside the shell \cite{Bell}. 
However these strong fields decrease with time as soon as the SNR expands. SNR turned
out to accelerate particles to higher energies during the shorter time at the very 
beginning of their history. It gives rise to the steepening of the energy spectrum 
and imitating the knee \cite{Drury,Ptus1}.

There are so called compact source models which are the combination of the source and 
interaction models. They examine the possibility for protons emitted by the source to 
lose the energy in photopion reactions and for nuclei - to disintegrate after the 
interaction with intense radiation fields around these compact sources
\cite{Tkacz,Candi}. Though the knee formed by these process is sharper than in the 
propagation models considered below (S $\approx$ 0.35 instead of 0.3) it is not 
sufficient to explain the observations. Moreover these models predict the extremely 
light mass composition beyond the knee, which contradicts the observations.    

I think all the mentioned source models have problems when compared with the shower 
phenomenology. Models which require different sources below and above the knee usually 
propose more energetic and less abundant sources of higher energy CR above the knee. 
They inevitably meet problems with the irregularity. The small number of high energy 
sources or shorter time of the acceleration to higher energies increase the sensitivity
of the cosmic ray energy spectrum to the particular space-time distribution of the 
sources and reduce the possibility to form the regular power-law spectrum. Source 
models need the additional mechanism (Halo or IGM) which could smooth the 
irregularity of the CR spectrum below and above the knee. 
On the other hand the models with SNR accelerating particles up to sub-EeV and EeV 
energies give just the smooth steepening of the CR energy spectrum and are not able to 
explain the sharpness of the knee.     

\subsubsection{Propagation models}

Propagation models were the first suggested after the discovery of the knee 
\cite{Peter}. They associate the steeper spectrum above the knee with the increased
leakage of CR particles from the Galaxy when their gyroradius becomes comparable with 
the transverse size of the Galactic Disk. Latest developments of this idea included 
models with introduction of the Hall diffusion \cite{Ptus2,Kalmy,Roule} and an 
anomalous diffusion in the ISM of quasi-fractal type \cite{Lagut}.  

All they have problems with the sharpness of the observed spectrum. Original models
had the sharpness of the constituent spectra not greater than 0.6. For the mixed mass 
composition with the abundance of constituent nuclei equal to their abundance at TeV 
energies the model sharpness decreased down to 0.3. In more sophisticated models with 
the Hall diffusion and a special geometry of regular magnetic field in the Galactic 
Disk and Halo the authors \cite{Ptus2,Kalmy} created the small intensity bump in the 
knee region and by this way increased the sharpness of constituent spectra up to 1.0.
The composite spectrum with a mixed composition will also have a greater sharpness but 
no more than 0.7. It is much less than the required value of about 3.

The second problem is the observed fine structure of the spectrum. If the propagation 
models predict the smooth constituent spectra with a continuous steepening, it is 
impossible to construct a total spectrum with discontinuities which are seen as
 the fine structure.

\subsection{Interaction models}

The interaction models introduce new processes in the interaction of CR particles 
at PeV energies to justify the sharp steepening of the EAS size spectrum. Some of 
these models require the new physics either to create new heavy particles with large 
cross-sections, which decay into yet unobservable high energy muons and neutrinos 
\cite{Petru,Dova}, some required emission of technihadrons, gravitons \cite{Kazan} 
or very many low energy particles \cite{Nikol}. The model with an enhanced production 
of charmed particles \cite{Yakov} slows down the development of the atmospheric cascade
 so that the bulk of its energy is dissipated not in the atmosphere but reaches the 
mountain or even the sea level and is absorbed in the earth.

The experimental confirmation of neutrino oscillations and therefore an existence of 
the neutrino mass inspired the creation of the models where cosmic ray particles 
interact with relic neutrinos in the gravitationally attracted neutrino halos 
surrounding galaxies and galaxy clusters \cite{Wigma}. By assuming the neutrino
mass in the region of eV it is easy to obtain the energy threshold for the inelastic 
$\nu$P interaction in the PeV region. 

In my mind the drawback of these models is not only that most of them require the new 
physics in the energy region very close to that already studied in Fermilab, but that 
they often don't specify the new process and don't develop their interaction model to 
incorporate and test it with well developed simulation codes such as CORSIKA, like it 
has been done for more conservative models (QGSJET, SYBILL etc.). These 
models can explain the existence of the knee, sometimes even its sharpness by 
introducing a rapid rise of the cross-section above the threshold, but never
other phenomenology of EAS accumulated hitherto. In this respect I should better 
attribute nearly all of them to ideas and not to models.               
          
\subsection{Phenomenological models}

Phenomenological models of the knee just extrapolate the observed phenomenological 
characteristics of CR or EAS into the high energy region and demonstrate that with 
certain approximations they give the steepening of the energy spectrum which they 
associate with the knee. 

To such type of models I attribute the model of Tsallis et al
\cite{Tsall}. They applied the formulae of generalized non-extensive statistics to the 
observed CR energy spectrum and have found that it is possible to fit its behaviour 
with the bend at sub-GeV and the steepening at PeV energies in the total range of 34 
decades of the differential intensity. The knee in this model appears as a crossover 
between two fractal thermal regimes.       

To the phenomenological class of models belongs also the multi-component model of 
Biermann and Ter-Antonyan \cite{Bier3}. They obtain parameters of the constituent
nuclei spectra by multivariate fitting of the electron and muon size spectra measured 
by KASCADE and MAKET ANI experiments.
      
Another phenomenological model is proposed by Stenkin \cite{Stenk}, who connects the 
steepening of the EAS size spectrum with the non-linearity of the $\log{N}_e(\log{E}_0)$ 
dependence which in turn is connected with the change of the EAS structure 
as soon as the shower maximum approaches the observation level. If $N_e \propto E^s$ 
where $s$ is the decreasing function of $E_0$ then even for the power law primary 
energy spectrum $I(E_0) \propto E_0^{-\gamma}$ with a constant $\gamma$ the slope index
$\alpha=\frac{\gamma}{s}$ of the EAS size spectrum $I(N_e) \propto N_e^{-\alpha}$ 
increases with $N_e$ and the spectrum steepens imitating the knee even for the constant
 slope of the primary spectrum.

I would attribute  to phenomenological models also the so-called 
{\it poly-gonato model} of J.R.H\"{o}randel \cite{Hoer2}. He noticed that spectra of 
constituent nuclei in primary 
CR become flatter with an increasing primary mass. He extrapolates these 
spectra up to their rigidity cutoff and found that it is possible to describe not only 
the knee but the general behaviour of the primary energy spectrum up to sub-EeV 
energies. At these high energies the substantial fraction of cosmic rays can be 
trans-iron elements. 

Since these models are purely phenomenological and not pretend to explain the origin
of cosmic rays they have no problems with the spectrum irregularities. However they 
may have problems with the observed fine structure of the spectrum and definitely 
with the sharpness of the spectrum if they will not assume the sharp cutoff of the 
constituent spectra. 
   
\section{Status of the Single Source Model}

It would be unfair if I'll tell nothing about the present status of the Single Source 
Model (SSM). Me and Prof.A.W.Wolfendale proposed it in 1997 \cite{EW1} and updated 
in 2001 \cite{EW2}. The main idea was that the knee is due to the contribution of just
one single, recent and nearby SNR. Since it is nearby and recent, CR from this SNR have
a flat energy spectrum with a sharp cutoff \cite{Berez}. At PeV energies their 
intensity becomes comparable with the total CR background from all other distant and 
old SNR and their contribution creates the bump which forms the knee. Due to the sharp 
cutoff of the spectrum beyong its maximum rigidity $R_{max}$ the knee is sharp and 
different constituent nuclei which have different maximum energy $E_{max} = ZR_{max}$ 
give rise to the fine structure of the spectrum.

Actually the energy spectrum of CR created by the Single Source is the result of the 
fit of theoretical spectra \cite{Berez} to the existing experimental data from 40 
independent EAS electron size spectra, 9 muon size spectra, 3 hadron size and energy
spectra and 5 Cherenkov light spectra. Comparing this spectrum with the spectrum of CR
expected from SNR of different ages and placed at different distances from the Earth we
 estimated the most likely intervals of distances and ages of the Single Source 
\cite{EW5}. They 
appeared to be rather narrow around 300 pc and 100 kyear. Soon after that american 
astronomers using the triangulation technique have found that the pulsar B0656+14 
associated with the SNR Monogem Ring is situated exactly at the indicated range of 
distances and ages as required for our Single Source \cite{Thors}.

In \cite{EW5} we presented arguments why the nearby SNR is not seen 
in gamma rays. It is because being nearby it is situated in our Local Superbubble with 
its low density of ISM which is about $3\cdot 10^{-3}$ cm$^{-3}$.
Moreover it is not a discrete, but an extended source with the angular radius of 
20$^\circ$ and the problem of the correct subtraction of the background from such a 
big area is not trivial. Many experiments 
have already looked for an excessive intensity of CR from this region and found nothing
 \cite{Cygn,Casa,Sun,Mila1,Airob,Tibet,Kasc1,Kasc2,Mila2,Chili}. Nevertheless the 
situation is controversial, since recently 
the Moscow University \cite{Zotov} and Tien-Shan \cite{Benko} groups revealed weak
 EAS excesses in the Monogem Ring area. If there are indeed CR excesses at sub-PeV and
 PeV energies from the Monogem Ring region they can evidence for the activity of the 
pulsar B0656+14 which emitted high energy CR soon after its birth and which were
confined within the SNR for a long time of about 80 kyear \cite{EW4}. High energy 
gamma-quanta which create these excesses are produced by CR on the local 
fluctuations of the interstellar gas density. 

The energy spectum of CR produced by the
pulsar B0656+14 has the sharp peak at the rigidity of 0.25 PV which is very close to 
the $R_{max} \approx 0.4$ PV, required by the SN model of Berezhko et al \cite{Berez} 
and can contribute to the observed knee sharpness, but its contribution to the CR 
intensity at the knee does not exceed 15\%. However, this pulsar can be responsible 
for CR beyond the knee at energies above 10 PeV. Recently we 
analysed the region beyond the knee and have found an extended excess of EAS intensity 
in the broad region slightly above $10^8$ GeV \cite{EW3} (Figure 4).  
 The existence of this excess gives an argument in favor of the 
assumption that the pulsar B0656+14 contributes to the CR intensity above $10^7$ GeV. 
So that there is an impression that the SNR Monogem Ring with its 
associated pulsar B0656+14 can be a serious contender for the Single Source 
responsible for the PeV CR intensity bump and the knee itself.
\begin{figure}
\centerline{
 \epsfig{file=                  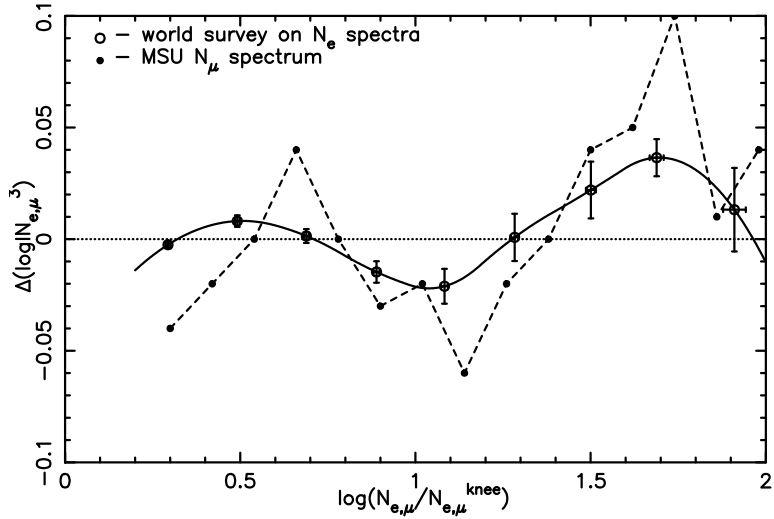,height=0.8\linewidth,angle=-90}
}
\vspace*{8pt}
\caption{The structure of the EAS size spectra beyond the knee (full line). 
Dashed line shows the EAS muon size spectrum measured in the Moscow University 
$^{50}$}
\end{figure}

It is not easy to kill the SSM, since its parameters were taken to fit the existing 
phenomenology. It is possible, however, to check and to kill the initial mass 
composition predicted by SSM. Originally we attributed intensity peaks at 3 and ~12
PeV to $O$ and $Fe$. If we are right then we cannot expect another peaks beyond 12 PeV.
However if the observed peaks are actually due to $He$ and $O$, then we can expect the 
existence of the sharp $Fe$ peak at 40 PeV. So far there is no indication of this peak 
in the experimental data from KASCADE. We look forward at the new data from 
KASCADE-Grande.

\section{Summary}

In the summary I stress again that the models which pretend to explain the
origin of the knee should explain not only the high isotropy, the tendency of the 
primary mass composition to become heavier above the knee, but also such 
phenomenological features as the {\it sharpness}, the {\it fine structure} of the knee 
and the relatively {\it small irregularity} of the energy spectrum both {\it far below}
and {\it far above} the knee. At the moment none of the models can satisfy these 
requirements. 

Source models which require less abundant and more energetic sources have
 problems with the large irregularity of the spectrum in particular above the knee. 
The same difficulty have the models with relatively abundant sources like pulsars which
 emit high energy CR for a short time after their creation. Definitely we have to 
incorporate the object or the mechanism which level out irregularities originated from 
the stochastic space-time distributions of sources. The first candidates for such an 
object could be the Galactic Halo or IGM.

All propagation models have not the sufficient sharpness of the knee even for the 
constituent spectra. As the consequence they should have problems with the observed
fine structure of the spectrum. They take for granted the regular power law spectra 
with no explanation of the origin of this regularity, therefore they can also have 
hidden problems with the irregularity of the spectrum.

The phenomenological models don't pretend to explain the origin of CR and the knee - 
they use different input spectra and find the best fit to the observed phenomenology
of the showers. The sharpness, the fine structure depend on the shape of these input 
spectra so that there is no actual explanation of the observed phenomenology.

The interaction models, besides that they require 'the new physics', are not 
sufficiently developed to be tested by the phenomenology. It is clear that modifying 
the threshold behaviour of the new process it is possible to fit both the sharpness 
and the fine structure of the spectrum, but the value of this exercise is not very
high. Since the main subject of interaction models is the knee they do not explain the 
small irregularity of the CR spectrum outside the knee region.

Schematically the possibility for different models to match the phenomenology of EAS 
in the knee region is outlined in Table I. 
\begin{table}[htb]
\tbl{Correspondence between models and the phenomenology}
{\begin{tabular}{@{}cccc@{}}                                       \toprule
Model/Phenomenology & Sharpness & Fine structure & Irregularity \\ \colrule
Source           & +  & + & ?             \\
Propagation      & -- & -- & not examined \\
Phenomenological & +  & +  & not examined \\
Interaction      & +  & +  & not examined \\                       \botrule
\end{tabular}}
\end{table}
This table shows that the {\it astrophysical} models are the most developed and some 
of them can match the observed phenomenology. Among astrophysical models {\it source} 
models are the most plausible. For nearly all models we need to introduce a mechanism
which could smear out the irregularity of the energy spectrum due to the stochastic nature
of the sources, viz.\ Galactic Halo, Intergalactic Medium or so.


\begin{thebibliography}{51}
\bibitem{Hoer1} J. R. H\"{o}randel, {\it Astropart.\ Phys.}, {\bf 21}, 241 (2004).
\bibitem{Kulik} G. V. Kulikov and G. B. Khristiansen, {\it JETP}, {\bf 35}, 635 (1958).
\bibitem{EW1}   A. D. Erlykin and A. W. Wolfendale, {\it J.\ Phys.\ G: Nucl.\ Part.\ Phys.},
                {\bf 23}, 979 (1997).
\bibitem{EW2}   A. D. Erlykin and A. W. Wolfendale, {\it J.\ Phys.\ G: Nucl.\ Part.\ Phys.},
                {\bf 27}, 1005 (2001).
\bibitem{Bier1} P. L. Biermann, {\it New Ast.\ Rev.}, {\bf 48}, 41 (2004); 
                astro-ph/0309810.
\bibitem{Bier2} P. L. Biermann et al., astro-ph/0302201 (2003). 
\bibitem{Svesh} L. G. Sveshnikova, {\it A\&A}, {\bf 409}, 799 (2003).
\bibitem{Dar1}  A. Dar, astro-ph/9809163 (1998).
\bibitem{Dar2}  A. Dar and R. Plaga, astro-ph/9902138 (1999).
\bibitem{Plaga} R. Plaga, {\it New Astronomy}, {\bf 7}, 317 (2002).
\bibitem{DeRu1} A. De Rujula, astro-ph/0207033 (2002).
\bibitem{Dar3}  A. Dar, astro-ph/0408310 (2004).
\bibitem{DeRu2} A. De Rujula, this conference.
\bibitem{Bell}  A. R. Bell and S. G. Lucek, {\it MNRAS}, {\bf 321}, 433 (2001).
\bibitem{Drury} L. 0'C. Drury et al., astro-ph/0309820 (2003).
\bibitem{Ptus1} V. S. Ptuskin and V. N. Zirakashvili, 2003, A\&A, {\bf 403}, 1; 
                astro-ph/0408025.
\bibitem{Tkacz} W. Tkaczyk {\it 27th ICRC, Hamburg}, {\bf 5}, 1979 (2001).
\bibitem{Candi} J. Candia et al., {\it Astropart.\ Phys.}, {\bf 17}, 23 (2002).
\bibitem{Peter} B. Peters, {\it 6th ICRC, Moscow}, {\bf 3}, 157 (1959).
\bibitem{Ptus2} V. S. Ptuskin et al., {\it A\&A}, {\bf 1}, 1 (1993).
\bibitem{Kalmy} N. N. Kalmykov and A. I. Pavlov, {\it 26th ICRC, Salt Lake City},
                {\bf 4}, 263 (1999).
\bibitem{Roule} E. Roulet, astro-ph/0310367.
\bibitem{Lagut} A. A. Lagutin and V. V. Uchaikin, {\it 27th ICRC, Hamburg}, {\bf 5},
                1900 (2001).
\bibitem{Petru} A. A. Petrukhin, {\it 27th ICRC, Hamburg}, {\bf 1}, 1 (2001);
                {\it Phys.\ Atomic Nucl.}, {\bf 66}, 517 (2003).
\bibitem{Dova}  M. T. Dova et al., astro-ph/0112191 (2001).
\bibitem{Kazan} D. Kazanas and A. Nikolaidis, {\it 27th ICRC, Hamburg}, {\bf 5}, 1760 
                (2201); astro-ph/0103147, hep-ph/0109247 (2001).
\bibitem{Nikol} S. I. Nikolsky, {\it Nucl.\ Phys.\ B (Proc.\ Suppl.)}, {\bf 39A}, 228
                (1995); {\it 27th ICRC, Hamburg}, {\bf 4}, 1389 (2001).
\bibitem{Yakov} V. I. Yakovlev, {\it 25th ICRC, Rome}, {\bf 1}, 446 (1995).
\bibitem{Wigma} R. Wigmans, {\it Astropart.\ Phys.}, {\bf 19}, 379 (2003).
\bibitem{Tsall} C. Tsallis et al., {\it Phys.\ Lett.\ A}, {\bf 310}, 372 (2003).
\bibitem{Bier3} P. L. Biermann and S. V. Ter-Antonyan, {\it 28th ICRC, Tsukuba}, {\bf 1}, 
                235 (2003).
\bibitem{Stenk} Yu. V. Stenkin, {\it 28th ICRC, Tsukuba}, {\bf 1}, 267 (2003); 
                {\it Mod.\ Phys.\ Lett.\ A}, {\bf 18}, 1225 (2003).
\bibitem{Hoer2} J. R. H\"{o}randel, {\it Astropart.\ Phys.}, {\bf 19}, 193 (2003).
\bibitem{Berez} E. G. Berezhko et al., {\it JETP}, {\bf 82}, 1 (1996).
\bibitem{EW5}   A. D. Erlykin and A. W. Wolfendale, {\it J.\ Phys.\ G: Nucl.\ Part.\ Phys.},
                {\bf 29}, 709 (2003).
\bibitem{Thors} S. Thorsett et al., {\it ApJ Lett.}, {\bf 592}, L71 (2003).
\bibitem{Cygn}  D. E. Alexandreas et al. {\it ApJ}, {\bf 383}, L53 (1991).
\bibitem{Casa}  T. McKay et al. {\it ApJ}, {\bf 417}, 742 (1993).
\bibitem{Sun}   L. Sun and S. Sun {\it 25th ICRC, Durban}, {\bf 4}, 165 (1997).
\bibitem{Mila1} K. Wang et al. {\it ApJ}, {\bf 558}, 477 (2001).
\bibitem{Airob} F. A. Aharonian et al. {\it A\&A}, {\bf 390}, 39 (2002).
\bibitem{Tibet} S. W. Cui et al. {28th ICRC, Tsukuba}, {\bf 4/7}, 2315 (2003).
\bibitem{Kasc1} T. Antoni et al. {\it ApJ}, {\bf 604}, 687 (2004).
\bibitem{Kasc2} T. Antoni et al., astro-ph/0402656 (2004).
\bibitem{Mila2} R. Atkins et al., astro-ph/0403097 (2004).
\bibitem{Chili} A. A. Chilingaryan et al., this conference.
\bibitem{Zotov} G. V. Kulikov and M. Yu. Zotov, 28th RCRC, Moscow, 2004; 
                astro-ph/0407138.
\bibitem{Benko} D. Benko et al., 28th RCRC, Moscow, 2004 (to be published).
\bibitem{EW4}   A. D. Erlykin and A. W. Wolfendale, {\it Astropart.\ Phys.}, {\bf 22/1},
                47 (2004).
\bibitem{EW3}   A. D. Erlykin and A. W. Wolfendale (in preparation).
\bibitem{Fomin} Yu. A. Fomin et al., {\it 28th ICRC, Tsukuba}, {\bf 1}, 119 (2003).
\end{thebibliography}
\end{document}